\documentclass[preprint,superscriptaddress,aps,pra]{revtex4-1}
\usepackage{amsfonts}
\usepackage{amsmath}
\usepackage{graphicx}

\emergencystretch=\maxdimen
\begin{document}

\title{Pulse Width Modulation method for quantum control design}
\author{Qi-Ming Chen}
\altaffiliation{Q.M.Chen is currently at the Department of Chemistry, Princeton University as a visiting scholar.}
\affiliation{Department of Automation, Tsinghua University, Beijing, 100084, China}
\affiliation{Center for Quantum information Science and Technology, Tsinghua National Laboratory for Information Science and Technology, Beijing 100084, China}
\author{Re-Bing Wu}
\email{rbwu@tsinghua.edu.cn}
\affiliation{Department of Automation, Tsinghua University, Beijing, 100084, China}
\affiliation{Center for Quantum information Science and Technology, Tsinghua National Laboratory for Information Science and Technology, Beijing 100084, China}
\author{Herschel Rabitz}
\affiliation{Department of Chemistry, Princeton University, Princeton, New Jersey 08544, USA}

\date{\today}

\begin{abstract}
	Illuminated by the Pulse Width Modulation (PWM) technology in classical control engineering, we propose the PWM approximation which transforms continuous and bang-bang control into each other. This method works by squeezing the approximation error into high-frequency components above a prescribed cutoff frequency $\Omega$, which can be tailored by modulating the switch time of the bang-bang control. Because the quantum dynamics under bang-bang control can be efficiently calculated, PWM allows us to speed up quantum control design particularly in large dimensional systems with small number of control variables. Moreover, PWM paves the way to implement various bang-bang control protocols in the literature with an arbitrary waveform, which is more desirable in laboratory. 
\end{abstract}

\maketitle
\section{Introduction}
Engineering of quantum systems is expected to lead to profound insights in physics as well as to novel applications \cite{Vandersypen2005}. Among all the control schemes, bang-bang and continuous controls are the two forms widely studied in quantum control systems. Quantum bang-bang control works by turning on/off the interactions in the system repeatedly without modifying the interaction strength \cite{Viola1998}, i.e., it utilizes all the control variables to the extreme if the switch is turned on, while leaves the system at rest (, or interacts with the uncontrolled environments) if turned off. Due to the simplicity of this method, it makes it possible to generate analytical results, and has been investigated in various theoretical studies, e.g., dynamical decoupling (DD) \cite{Viola1998, Viola1999, Viola1999a, Viola2003}, cavity state engineering \cite{Law1996, Liu2004}, and quantum time-optimal control (TOC) \cite{Khaneja2001}.

Though the bang-bang control is popular in theoretical researches, the fast switch required is not always available in real experiments. Superconducting circuit, for example, requires the control field to be slow enough to avoid undesired coupling with extra energy levels \cite{Blais2004, Blais2007}. Quantum continuous control works by dynamically modifying the phase and amplitude of the control fields, which is more desirable in a laboratory \cite{Sundermann1999}. Successful applications of quantum continuous control include controlled adiabatic passage (STIRAP) \cite{Vitanov2017}, quantum optimal control (QOC) \cite{Peirce1988, Warren1993, Assion1998, Khaneja2005}, and quantum simulation of time-dependent Hamiltonians\cite{Berry2007, *Wiebe2010, Poulin2011}.

A natural question arises: whether these two types of controls can be transformed from each other. If does, one may utilize the simplicity of bang-bang control in theoretical analysis, but conduct experiments with a continuous control field. Or, one may design various algorithms to speed up quantum control design \cite{Yip2003}. Fortunately, Pulse Width Modulation (PWM) constructs a bridge between the two controls. PWM is a typical technology in classical control systems, which has a very special feature that the control function is physically bang-bang but effectively continuous \cite{Holmes2003,  Sun2012}. Given a continuous function, one can always find a corresponding bang-bang control sequence, i.e., the PWM sequence, which approximates the continuous control function at any precision. In other words, by investigating the PWM method in quantum control systems we obtain a new control method which embeds the advantages of both the bang-bang and continuous controls. In addition, we may transform the various bang-bang control results in the literature into a more experimental feasible form, i.e., the continuous control.

The rest of this paper is organized as follows. In Sec. II, we introduce the PWM approximation which approximate an arbitrary control field with a sequence of bang-bang pulses. Analyses show that the sequence controls the system with $2$nd order accuracy compared with the original control field, which can be more accurate when pulses are concatenated in a particular way. In Sec. III, we show that PWM method speeds up the numerical calculation in solving time-dependent Schr\"odinger equation, especially in large dimensional systems with a small number of control variables. Computational complexity of several propagation methods are also compared in this section. To realize the designed bang-bang control sequence, or transform bang-bang control results in the literature into a more useable form, Sec. IV proposes methods which transform the designed bang-bang control  into arbitrary waveforms or a \emph{Gaussian} pulse train. Sec. V presents an application of the PWM method in molecular systems, i.e., combine PWM and GRAPE algorithm to speed up quantum control design in a ten-dimensional system. The results are accord with theoretical analyses. We draw conclusions and discuss further applications of PWM in Sec. VI.

\section{The PWM approximation of arbitrary waveforms}
\begin{figure}
  \centering
  \includegraphics[width=8cm]{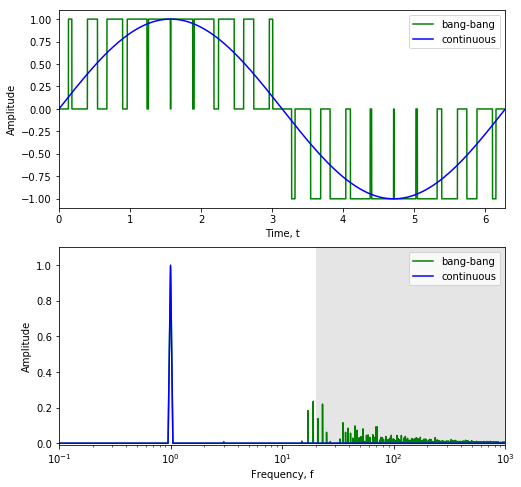}
  \caption{(Color online) The PWM approximation (green) of a sinusoidal function (blue) with cutoff frequency $\Omega=19$Hz. In each period $T=2\pi$, the sinusoidal function is approximated by $M=20$ pulses, where the pulse amplitude is chosen as $\xi=1$, $\tau=T/M$. Fourier analysis shows that the two fields are only different above the cutoff frequency (dark area).}\label{REC}	
\end{figure}

Consider an arbitrary time-dependent Hamiltonian 
\begin{equation}\label{H_T}
  H(t) = H_0+\sum_{k=1}^{K}u_{k}(t) H_{k},
\end{equation}
where $H_0$ represents the constant part and the summation the time-varying parts. Real functions $u_k(t)$'s are tunable driving fields applied on the system, which are assumed to have a finite bandwidth $[\omega_{\rm min}, \omega_{\rm max}]$ \cite{Vandersypen2005}. 

Define $\Omega \gg \omega_{\rm max}$ as the cutoff frequency of the control fields chosen according to the desired approximation precision, the aim of PWM approximation is to replace each of the continuous control fields $u_k(t)$ by sequence of rectangular pulses (bang-bang control) $s_k(t)$, where the approximation error lies only above the threshold in the frequency domain. The PWM sequence is defined by splitting every time interval $T=2\pi/\omega_{\rm min}$ into $M=\Omega/\omega_{\rm min}+1$ subintervals with equal length $\tau$, within which there exists one rectangular pulse located at the center, i.e.,  
\begin{equation}\label{R_PULSE}
	s_k(t) = \left\{
                  \begin{array}{ll}
                    \xi_k \cdot sgn\Big\{w_k^{(m)}\Big\}, 
                    & \hbox{$\left| t-t^{(m)}\right| \leq \left| w^{(m)}\right|/2$} \\
                    0,
                    & \hbox{else}
                  \end{array}
                \right.
\end{equation}
where $t^{(m)}$ is the center of every subinterval, $\xi_k$ is the common amplitude of all the pulses, $|w_k^{(m)}|$ is the pulse width of the $m$-th pulse, satisfying
\begin{equation}\label{WID}
	w_k^{(m)}=\xi_k^{-1}\int_{t_{-}^{(m)}}^{t_{+}^{(m)}}dt'u_k(t'),
\end{equation}
where $t_{-}^{(m)}=t^{(m)} - \tau/2$, $t_{+}^{(m)}=t^{(m)} + \tau/2 $ (see Appendix \ref{APP} for detail). This formula indicates that the PWM sequence is only determined by the integral area of the time-dependent function $u(t)$ in every small time interval Figure \ref{REC} compares the two types of control fields $u_k(t)$ and $s_k(t)$ in both time domain and frequency domain. Although they look rather different in the time domain, i.e., one is bang-bang but the other is continuous, their \emph{Fourier} transform are identical below the prescribed cutoff frequency $\Omega$. In other words, the PWM approximation squeezes errors into the high-frequency region above $\Omega$.
\begin{figure}
  \centering
  \includegraphics[width=8.5cm]{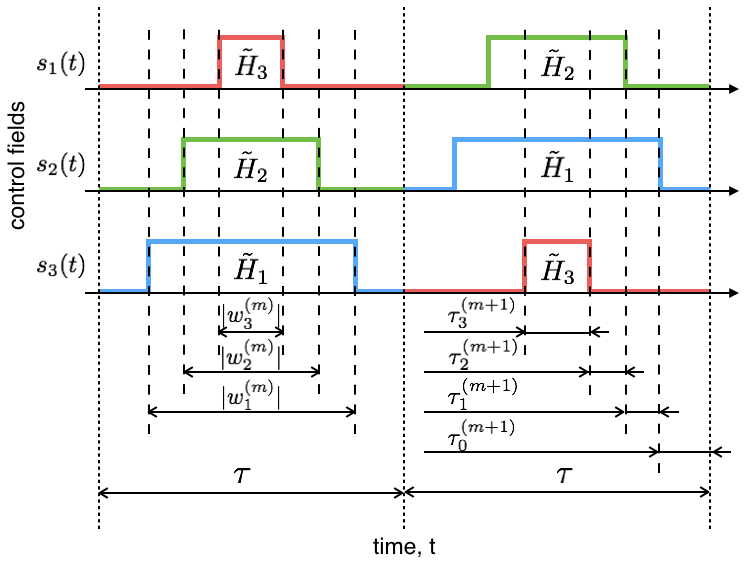}
  \caption{(Color online) Schematic of the PWM control process. Parameters $w_k^{(m)}$'s and $\tilde{H}_k$'s are labeled according to the pulse width, i.e., the blue rectangles always correspond to the label $k=1$, green to $k=2$, and red to $k=3$, but with no explicit relation to $H_k$. }\label{SCH}
\end{figure}

By replacing all the control variables by pulse sequences, the time-dependent Hamiltonian $H(t)$ is approximated by a sequence of time-independent Hamiltonians with prescribed cutoff frequency $\Omega$, i.e.,  
\begin{equation}\label{H_PWM}
	H_{\rm PWM}(t) =  H_0+\sum_{k=1}^{K}s_{k}(t) H_{k}.
\end{equation}
Since $s_k(t)$'s are all bang-bang control functions defined in Eq. (\ref{R_PULSE}), $H_{\rm PWM}(t)$ at any instinct $t$ is chosen from the following finite set
\begin{eqnarray}
	\mathcal{S}=\Big\{ H_0 + \sum_{k=1}^{K}\delta_k\cdot\xi_k H_k | \delta_k = 0, \pm 1 \Big\}.
\end{eqnarray} 
Figure \ref{SCH} illustrates the schematic of the approximated Hamiltonian sequence described in Eq. (\ref{H_PWM}). In every small subinterval $\tau$, it requires the interactions with the system  to be turned on in turns, waiting for time duration $|w_k^{(m)}|$ respectively before turned off. For a Hamiltonian with $K$ independent time-dependent variables, it needs $2K$ times of switch in total over every subinterval. 

To evaluate the precision of the PWM approximation, we assume that in the $m$-th subinterval
\begin{equation}
\tau \geq \left|w_{k_1}^{(m)}\right| \geq \cdots \geq \left|w_{k_K}^{(m)}\right|,
\end{equation}
and define $\tilde{H}_n=sgn(t_{k_n}^{m})\cdot \xi_{k_n}H_{k_n},~n=1,\cdots,M$. According to the Hamiltonian described in Eq. (\ref{H_PWM}), time propagation of the system can be written as
\begin{eqnarray}\label{UP}
  \nonumber U_{\rm{PWM}}^{(m)}&&=\exp{\left[ -i\tau_{0}^{(m)} \tilde{H}_0 \right]}
  \exp{\left[ -i\tau_{1}^{(m)}\left(\tilde{H}_0 + \tilde{H}_1\right) \right]}\times \cdots
  \times
  \exp{\left[ -i\tau_{K}^{(m)} \left(\tilde{H}_0 + \cdots + \tilde{H}_K\right) \right]}\\
  &&\times \cdots \times
  \exp{\left[ -i\tau_{1}^{(m)}\left(\tilde{H}_0 + \tilde{H}_1\right) \right]}
  \exp{\left[ -i\tau_{0}^{(m)}\tilde{H}_0 \right]},
\end{eqnarray}
where 
{\small \begin{equation}
	\tau_k^{(m)}=\left\{
                  \begin{array}{lll}
                   1/2\cdot\left( \tau - \left|w^{(m)}_{1}\right| \right),
                    & \hbox{$k=0$} \\
                   1/2\cdot\left( \left|w^{(m)}_{k}\right| - \left|w^{(m)}_{k+1}\right| \right),
                    & \hbox{$k=1,\cdots, {K-1}$} \\
                   \left|w^{(m)}_{K}\right|,
                    & \hbox{$k=K$}
                  \end{array}
                \right.
\end{equation}}
Compare Eq. (\ref{UP}) with the exact progator
\begin{equation}\label{UACC}
	U^{(m)}=\mathcal{T}\exp\left(\int_{t_-^{(m)}}^{t_+^{(m)}} dt H(t) \right),
\end{equation} 
where $\mathcal{T}$ represents time-order integral, we prove in Appendix \ref{APP2} that the PWM method approximates the short-time evolution with 2nd-order accuracy, i.e.,  
\begin{equation}\label{UACC}
	U^{(m)}=U_{\rm PWM}^{(m)} + \mathcal{O}(\tau^3).
\end{equation} 

If we choose $\xi_k\rightarrow\infty$, the PWM approximation becomes the Split Operator Scheme (SPO) \cite{Fleck1976, *Feit1982, *Feit1983, Bandrauk1992, *Bandrauk1993} (, also called the \emph{Suzuki-Trotter} Scheme \cite{Suzuki1985, *Suzuki1986, *Suzuki1990, *Suzuki1991, *Suzuki1995, Berry2007, *Wiebe2010, Poulin2011}). Thus, SPO can be seen as an special case of PWM. Compared with SPO, the PWM scheme saves half of the switching times but shows a higher approximation precision in approximating the exact time-evolution (see Appendix \ref{A_CPR}). Follow the same way in SPO, we prove in Appendix \ref{APP3} that the PWM pulse sequence can be concatenated to a form with arbitrarily higher-order precision. In particular, the $2n$th order PWM approximation can be written as
\begin{eqnarray}
	S_{2n}^{(m)}[\tau]&=& S_{2n-2}^{(m)}[s\tau] S_{2n-2}^{(m)}[(1-2s)\tau]
  S_{2n-2}^{(m)}[s\tau], \\
  S_2^{(m)}[\tau] &=& U_{\rm PWM}^{(m)},
\end{eqnarray}
where $s = {1}/{\left[ 2 + (-2)^{1/(2n-1)} \right]}$.

\section{Computational complexity of the PWM method}
\begin{figure}
  \centering
  \includegraphics[width=8cm]{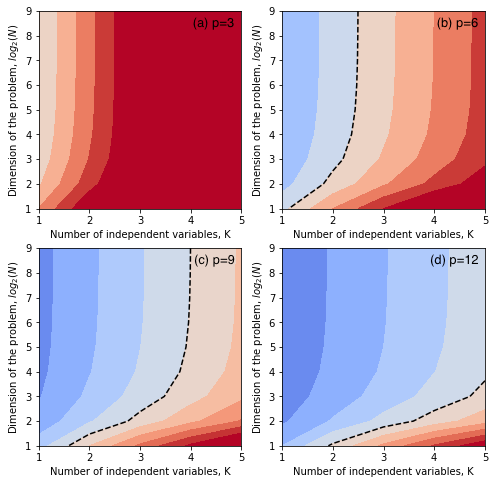}
  \caption{(Color online) Computational complexity compared between the PWM method and traditional method, where the color corresponds to the ratio $\gamma=\mathcal{C}_{\rm PWM}/\mathcal{C}_{\rm PWC}$. The blue area corresponds to situations where the PWM method is more efficient than the traditional method ($\gamma<1$), whereas the red area corresponds to situations where it is less efficient than the traditional method ($\gamma>1$). The black dashed line shows the boundary where the two methods are equally efficient ($\gamma=1$). }\label{CPX}
\end{figure}

Design of the quantum control fields needs efficient algorithms in solving time-dependent Schr\"odinger equation (TDSE). In \emph{Toolkit} method, Yip \emph{et al.} approximates the amplitude of a continuous control function $u(t)$ by $2n+1$ discrete values \cite{Yip2003}, in which the larger the parameter $n$ the more accurate the approximation  would be. However, parameter $n$ should not be too large in order to accelerate numerical simulation, so that there is a trade-off between approximation precision and computational complexity. Similar to that, the PWM method  approximates a continuous control field $u(t)$ by three discrete values, i.e., bang-bang control which corresponds to the lowest $n=1$ in \emph{Toolkit} method. To compensate the loss of control freedom, i.e., the field amplitude, PWM adjusts the short time intervals $w^{(m)}$ which guarantees 2nd order approximation accuracy in solving TDSE.

 but still guarantees the $2$nd order approximation precision in solving TDSE. 

Different from the time-dependent Hamiltonian described by Eq. (\ref{H_T}), $H_{\rm PWM}(t)$ is switched among only a finite number of time-independent Hamiltonians. Thus, one can diagonalize them and store the results before calculating the propagator, i.e.,  
\begin{equation}
  \tilde{H}_0 + \cdots \tilde{H}_k = D_{k} \Lambda_{k} D_{k}^{\dagger},~
  k=0,\cdots,K,
\end{equation}
where $\Lambda_k$'s are diagonal operators. Then, propagator presented in Eq. (\ref{UP}) can be written as
\begin{eqnarray}\label{UP_N}
  \nonumber U_{\rm{PWM}}^{(m)}
  &&=D_{0} \exp{\left[ -i\tau_{0}^{(m)} \Lambda_0 \right]}D_{0}^{\dagger}
  \times \cdots \times D_{K}
  \exp{\left[ -i\tau_{K}^{(m)} \left(\Lambda_0 + \cdots + \Lambda_K\right) \right]}D_{K}^{\dagger} \\
  &&\times \cdots \times
  D_{0} \exp{\left[ -i\tau_{0}^{(m)} \Lambda_0 \right]}D_{0}^{\dagger},
\end{eqnarray}

Since multiplication of variables usually requires more operations than addition on a computer, we define computational complexity $\mathcal{C}$ as the number of multiplications required when calculating short-time evolution. In the literature, short-time propagation methods are usually referred to the Piecewise Constant Scheme (PWC) \cite{Kosloff1988}
\begin{equation}
	U_{\rm PWC}^{(m)}=\exp\left[ -i\tau H(t')\right],~t'\in[t_{-}^{(m)},t_{+}^{(m)}], 
\end{equation}
and SPO \cite{Fleck1976, *Feit1982, *Feit1983, Bandrauk1992, *Bandrauk1993, Suzuki1985, *Suzuki1986, *Suzuki1990, *Suzuki1991, *Suzuki1995}
\begin{equation}\label{SPO}
	U_{\rm{SPO}}^{(m)}=\prod_{k=0}^{K}\prod_{k=K}^{0}\exp\left[ -i\tau/2 H_k \right].
\end{equation}
We prove in Appendix \ref{A_CPR} that PWC, PWM, and SPO have the same order of accuracy in numerical simulation (, $2$nd order accuracy), but the latter two schemes usually have a much smaller computational complexity than the former one in high-precision computation. Figure \ref{CPX} shows the comparison of computational complexity between PWC and PWM methods (, the computational complexity of SPO is the same as PWM), where $p$ is defined as the order of accuracy in calculating the exponential function, i.e., $\exp[x]=1+x+\cdots+\mathcal{O}(x^{p+1})$. Note that the total accuracy of a scheme in each subinterval $\tau$ contains the approximation precision of the propagator $\varepsilon_1\approx 1-\tau^3$ and the numerical precision of the exponential function $\varepsilon_2\approx 1-\tau^{p+1}$, parameter $p$ should be sufficiently large to guarantee an accurate propagation with precision $(\varepsilon_1\varepsilon_2)^M$ in the whole time interval $T=M\tau$. Numerical simulation shows that PWM outperforms PWC when $p$ is relatively large, particularly in systems with large dimension but small number of control variables.

\section{Continuous realization of the designed bang-bang control}
\subsection{Arbitrary waveform generator}
In laboratory, a smooth control field is usually more desirable due to the restriction of Arbitrary Waveform Generator (AWG) on the time-domain response. Suppose that we have found a suitable PWM sequence $s_k(t)$, one way to transform it back to a continuous form is to apply the PWM approximation backward 
\begin{equation}\label{A_PWM}
	\int_{t_{-}^{(m)}}^{t_{+}^{(m)}}dt u_k(t) = \xi_k w_k^{(m)},~m=1,\cdots,M.
\end{equation}
According to \emph{Lagrange Mean Value Theorem}, we know that there exist at least one point $t'$ in the interval $[t_{-}^{(m)}, t_{+}^{(m)}]$ such that $u_k(t')=(1/\tau)\cdot\xi_k w_k^{(m)}$. Chosen these $t'$'s as known points, continuous controls $u_k(t)$'s with arbitrary order of smoothness can be generated through various interpolation methods \cite{Stoer1993}. Suppose the $u_k(t)$'s vary slowly in every small interval $\tau$, we can simply obtain the following PWC control fields that is well accepted in the quantum control literature 
\begin{equation}
	u_k(t)=(1/\tau)\cdot\xi_k w_k^{(m)},~t\in[t_{-}^{(m)}, t_{+}^{(m)}].
\end{equation}

Though the inverse-PWM approximation seems obvious with the PWM pulse sequences, it needs more attention when dealing with an arbitrary bang-bang control field. Suppose the minimum and the cutoff frequency of a bang-bang control is $\omega_{\rm min}$ and $\Omega$ respectively, to generate a corresponding continuous field one needs to divide every time interval $T=2\pi/\omega_{\rm min}$ into $M\geq\Omega/\omega_{\rm min}+1$ pieces and apply Eq. (\ref{A_PWM})  in every subinterval $\tau$.  In principle, the smaller the pieces divided, the more accurate the continuous control field would be, and thus any bang-bang control field can be approximated with a continuous waveform. However, when the cutoff frequency is rather large, or when the spectrum of the bang-bang control is too wide, this inverse-PWM approximation fails since the obtained continuous function also contains rapid ascent/descent in the time domain which cannot be realized in AWG. 

Another way to transform arbitrary bang-bang control field into continuous form is to use \emph{Fourier} transform. Suppose the cutoff frequency of a bang-bang control is $\Omega$, we can simply drop off those high-frequency terms above the threshold through a numerical/physical frequency filter. In fact, the above two methods of generating arbitrary waveform are quite similar. The only difference lies in the high-frequency terms above $\Omega$, where the former method allows small components in the area. In this regard, the latter method may result into a more smooth control field than the former one.

\subsection{Gaussian pulse train generator}
\begin{figure}
	\centering
  \includegraphics[width=8cm]{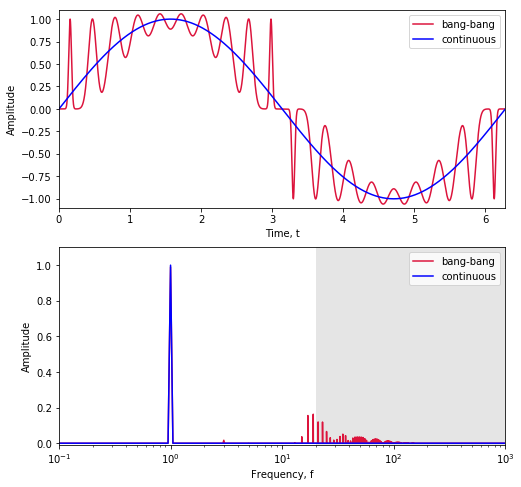}
  \caption{(Color online) The Gaussian pulse approximation (red) of a sinusoidal function (blue) with cutoff frequency $\Omega=19$Hz. In each period $T=2\pi$, the sinusoidal function is approximated by $M=20$ pulses, where the pulse amplitude is chosen as $\xi=1$, $\tau=T/M$. Fourier analysis shows that the two fields are only different above the cutoff frequency (dark area).}\label{GAU}
\end{figure}

Instead of transforming the designed PWM sequence into arbitrary waveform, we can also replace the rectangular pulses by a \emph{Gaussian} pulse train. Under the same definition of the parameters in Sec. II (A), we define the \emph{Gaussian} pulse sequence as
\begin{equation}
	s_k(t)=\xi_k e^{{-\pi \left(t-t^{(m)}\right)^2}/{\left(w_k^{(m)}\right)^2}},~ 
	t\in[0,T].
\end{equation}
Following the same way as in Appendix \ref{APP}, we can prove that the \emph{Gaussian} pulse sequence is identical to the PWM sequence below the cutoff frequency $\Omega$ in the frequency domain. Above the threshold, there are only small differences between the two types of sequences. 

For an arbitrary bang-bang control field with minimum and cutoff frequency $\omega_{\rm min}$ and $\Omega$, respectively, one needs to generate $M\geq\Omega/\omega_{\rm min}+1$ \emph{Gaussian} pulses in every time interval $T=2\pi/\omega_{\rm min}$, where the pulse width $w^{(m)}$ is defined in Eq. (\ref{WID}). Though hard switches are not available in experiments, a very narrow \emph{Gaussian} pulse is regarded to be achievable. In this regard, any bang-bang control can be well implemented with a \emph{Gaussian} pulse train. Figure \ref{GAU} compares the two types of control fields in both time domain and frequency domain. The approximation error is squeezed in the area above the cutoff frequency $\Omega$ in the frequency domain. Compared with the results in Fig. \ref{REC}, the \emph{Gaussian} pulse sequence is indeed a more accurate approximation to a continuous control function than the rectangular PWM sequence.

\section{Application: optimal control of molecular systems}
In ultra-fast laser control of molecular systems, the Hamiltonian is usually written as $H(t)=H_0 - \mu\epsilon(t)$ , where $H_0$ is the field-free Hamiltonian, $\epsilon(t)$ the control field which oscillates within a bounded range $\epsilon_{\rm min} \leq \epsilon(t) \leq \epsilon_{\rm max}$, and $\mu$ the dipole moment \cite{Yip2003}. Particularly in our example, we consider a ten-level system which is initially in the ground state $|\psi_i\rangle=|1\rangle$, our aim is to design the control field $\epsilon(t)$ which drives the system into the target state $|\psi_f\rangle=|4\rangle$ at time $T=100$ (, dimensionless units are used) \footnote{The ten-level quantum system in the illustration had the following properties, in dimensionless units. The energy levels, i.e., the diagonal elements of $H_0$) are $1$, $5$, $7$, $8$, $9$, $10$, $11$, $11.8$, $12.1$, and $12.4$. The elements of the dipole matrix are $\mu_{12}=0.3$, $\mu_{13}=0.15$, $\mu_{14}=0.0$, $\mu_{17}=0.003$, $\mu_{23}=0.2$, $\mu_{24}=0.25$, $\mu_{34}=0.1$ and all other non-diagonal elements are $0.001$.}. Thus, the objective function reads
\begin{equation}
	J = 1- \left| \langle \psi_f | U_T | \psi_i \rangle \right|^2 ,
\end{equation}
which corresponds to optimizing the overlap between the target state and the final state of the system \cite{Schirmer2011}, $U_{T}$ is the propagation operator starts from $t=0$ to $t=T$. Setting $\tau=0.1$ and applying PWM approximation on the initial control field $\epsilon(t)$, the gradient of the propagator reads
 \begin{eqnarray}
 	\frac{\partial U_T}{\partial w_k^{(m)}} &=& -\frac{i}{2}U_M \big[ 
	U_{m}^{\dagger} V_k^{(m)} \tilde{H}_k \left( V_k^{(m)} \right)^{\dagger} U_{m}
	+ U_{m-1}^{\dagger} \left(W_k^{(m)}\right)^{\dagger}\tilde{H}_k W_k^{(m)} U_{m-1} \big],
 \end{eqnarray}
where
\begin{eqnarray}
	U_m &=& U_{\rm PWM}^{(m)} \cdots U_{\rm PWM}^{(1)}, ~U_T = U_M,\\
	U_{\rm PWM}^{(m)} &=& V_{K}^{(m)} D_K \exp\left[ -i\tau_K^{(m)} \Lambda_K \right] D_K^{\dagger}
	W_{K}^{(m)}, 
\end{eqnarray}
\begin{eqnarray}
	V_k^{(m)} &=& \prod_{j=0}^{k-1} D_j \exp\left[ -i\tau_j^{(m)} \Lambda_j \right] D_j^{\dagger}, \\
	W_k^{(m)} &=& \prod_{j=k-1}^{0} D_j \exp\left[ -i\tau_j^{(m)} \Lambda_j \right] D_j^{\dagger}.
\end{eqnarray}
 Thus, the gradient of the objective function $J$ reads 
{\small \begin{equation}\label{GRAD}
	\frac{\partial J}{\partial w_k^{(m)}} 
	= 2{\rm Re}\left\{ \langle \psi_i | U_M^{\dagger} | \psi_f \rangle 
	\langle \psi_f | \frac{\partial U_T}{\partial w_k^{(m)}} | \psi_i \rangle \right\}.
\end{equation}}

\begin{figure}
  \centering
  \includegraphics[width=8.5cm]{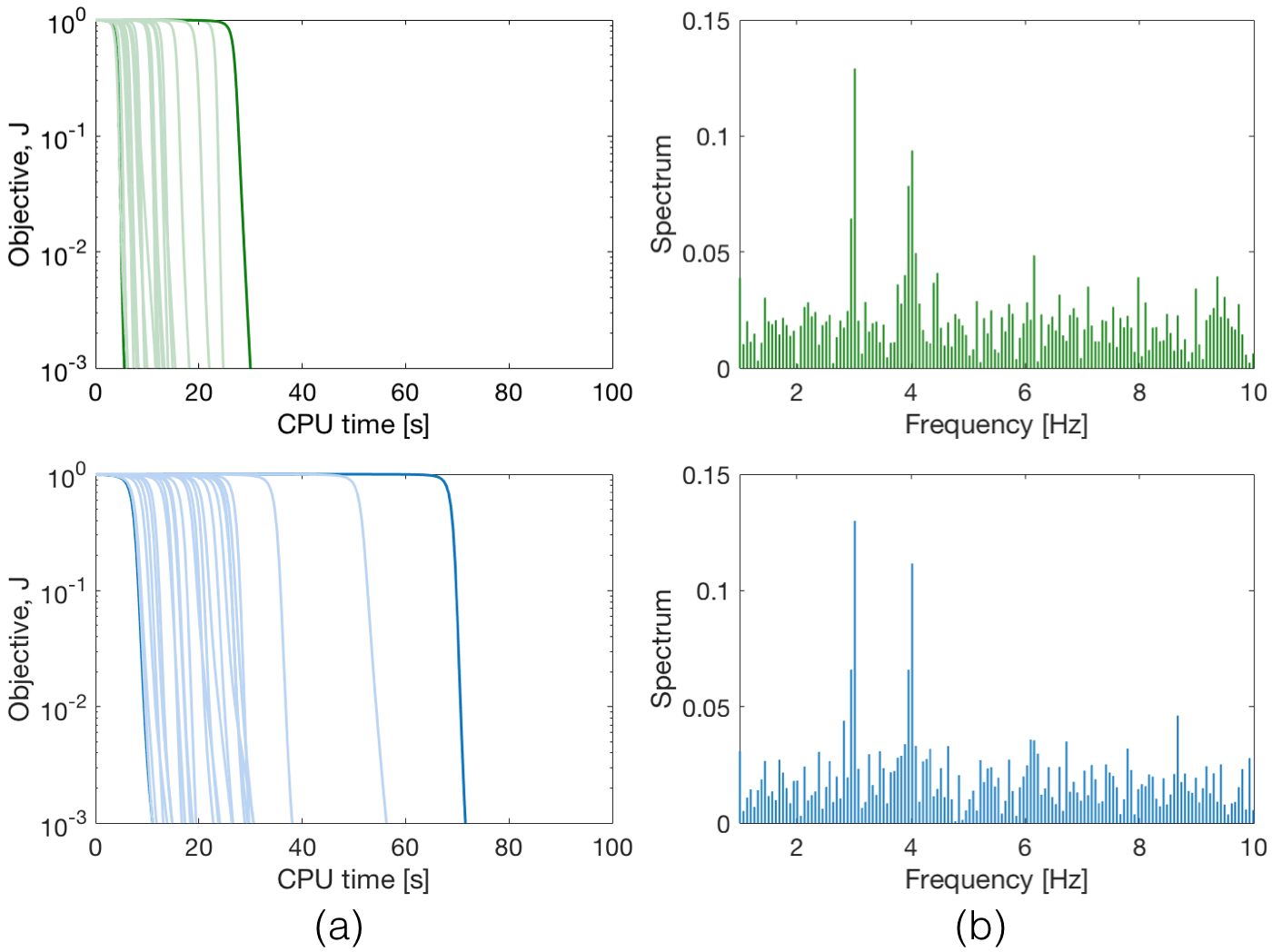}
  \caption{(Color online) (a) The search for the optimal control with error threshold $J\leq1\times10^{-3}$ by the PWM-GRAPE (green) and the basic GRAPE (blue) methods, and (b) the corresponding optimal controls. The two algorithms have been operated $25$ times each, where $T=100$, $\tau=0.1$, and the initial control field $\epsilon(t)$ is randomly chosen from $[-0.5,0.5]$.}\label{F_GRAD}
\end{figure}
Figure \ref{F_GRAD} (a) shows the search for the optimal control through the basic GRAPE algorithm with/without employment of the PWM method \cite{Khaneja2005}. In accordance with the analyses above, the PWM method diagonalize the three time-independent Hamiltonians $H_0$, $H_0 \pm \mu$ before solving TDSE, and thus accelerates the optimal control design by a large extent. On average, the PWM-GRAPE algorithm approximately saves more than half of the CPU time compared with the basic GRAPE algorithm, but still guarantees a high-fidelity control design.

The PWM-GRAPE algorithm involves an additional step to transform the optimized pulse sequence into continuous form. According to Eq. (\ref{A_PWM}), this can be done by various interpolation methods. In accordance with the basic GRAPE algorithm (, and almost other algorithms), we simply transform the optimized control field into PWC form $\epsilon(t)=(\xi/\tau)\cdot w^{(m)},~t\in[t_{-}^{(m)}, t_{+}^{(m)}]$. Figure \ref{F_GRAD} (b) shows the spectrum of the PWC control fields obtained by the PWM-GRAPE (green) algorithm and the basic GRAPE algorithm (blue), where both of the fields control the system with fidelity $J\leq 1\times 10^{-3}$. The optimized results show that the two coherent processes, i.e., $|1\rangle\rightarrow|2\rangle$ and $|2\rangle\rightarrow|4\rangle$, are dominant in the whole control process, which meets the results in Ref. \cite{Yip2003}.


\section{Conclusions}
To summarize, we propose the PWM method in quantum control design, which builds a connection between bang-bang and continuous control. On the one hand, this method approximates an arbitrary control field by bang-bang control, takes advantage of the discreteness of the control field to speed up numerical calculation when solving TDSE, or takes advantage of the higher-order propagation scheme to improve the numerical precision in the control design, and then approximates the designed bang-bang control by an arbitrary waveform or a \emph{Gaussian} pulse train. Seeing quantum optimal control in molecular systems as an example, we demonstrate that the PWM method can be embedded into the GRAPE algorithm (, almost any QOC algorithms) and accelerate the control design by a large extent. Both theoretical analysis and numerical simulation show that the it speeds up quantum control design especially in large dimensional systems with small number of control variables. On the other hand, the PWM method reveals one way to transform existing bang-bang control results in the literature into a laboratory available continuous form, or a \emph{Gaussian} pulse train. Our next step would be addressed on transforming bang-bang controls in TOC into arbitrary waveforms.

\appendix
\section{Derivation of the PWM approximation}\label{APP}
Consider a sequence of rectangular pulses $s(t)$ in which every single pulse is centered at $t^{(m)}=(m-1/2)\tau,~m=1,\cdots,M$, it can be \emph{Fourier} expanded as follows
\begin{eqnarray}\label{A_SUM}
  s(t)&&=\frac{\xi}{T}\sum_{m=1}^{M}w^{(m)}
  +\sum_{n\neq0}\Big[\sum_{m=1}^{M}\frac{\xi}{n\pi}
  \sin{\Big(\frac{w^{(m)}}{\tau}\frac{n\pi}{M}\Big)}
  e^{-in\omega t^{(m)}}\Big]e^{in\omega t},
\end{eqnarray}
where $M=T/\tau$ is the number of pulses in every time interval $T$, $\xi$ is the common amplitude of the pulses, $\left|w^{(m)}\right|\leq \tau$ is the pulse width for the $m$-th pulse in the sequence.

Suppose $M$ is a sufficiently large number, the second term of Eq. (\ref{A_SUM}) can be approximated with $\sum_{n\neq0}\Big[\sum_{m=1}^{M}\frac{\xi w^{(m)}}{M \tau}e^{-in\omega t^{(m)}}\Big]e^{in\omega t}$ when $n \ll M$. For those $n \approx M$ or $n \gg M$, the approximation above is also reasonable since the coefficient $1/n$ must be very small. Thus, Eq. (\ref{A_SUM}) can be simply written as
\begin{equation}\label{A_SP}
  s(t)\approx\frac{\xi}{T}\sum_{m=1}^{M}w^{(m)}
  +\sum_{n\neq0}\Big[\sum_{m=1}^{M}\frac{\xi w^{(m)}}{M \tau}e^{-in\omega t^{(m)}}\Big]e^{in\omega t}.
\end{equation}

If we choose
\begin{equation}\label{TP1}
  w^{(m)}=\frac{1}{\xi}\int_{t_{-}^{(m)}}^{t_{+}^{(m)}}dt\sin{(\omega t+\phi)},~T=2\pi/\omega, 
\end{equation}
where $t_{\pm}^{(m)}=t^{(m)}\pm \tau/2$, the first term in Eq. (\ref{A_SP}) can be eliminated
\begin{equation}\label{N0}
  \frac{\xi}{T}\sum_{m=1}^{M}w^{(m)}=0.
\end{equation}
For $n=1$, the summation over $m$ in the second term can be written as
\begin{eqnarray}
  \nonumber &&\sum_{m=1}^{M}\frac{\xi w^{(m)}}{T} e^{-i\omega t^{(m)}}e^{i\omega t}\\
  \nonumber &=&\frac{e^{i\omega t}}{4\pi}\Big\{ e^{i\phi}\sum_{m=1}^{M}
  \left(e^{-i\pi\frac{\tau}{T}}-e^{i\pi\frac{\tau}{T}}\right)
  +e^{-i\phi}\sum_{m=1}^{M}
  \left[e^{-i2\omega t^{(m)}}\left(e^{i\pi\frac{\tau}{T}}-e^{-i\pi\frac{\tau}{T}}\right)\right]\Big\} \\
  &\approx& \frac{e^{i(\omega t + \phi)}}{2i} 
  +i \frac{e^{i(\omega t - \phi)}}{4\pi}\sum_{m=1}^{M}e^{-i4\pi \frac{t^{(m)}}{T}}\cdot2\pi\frac{\tau}{T}\label{A1}\\
  &\approx& \frac{e^{i(\omega t + \phi)}}{2i} 
  +i \frac{e^{i(\omega t - \phi)}}{4\pi}\cdot\int_{0}^{2\pi}d\theta e^{-i2\theta}\label{A2}\\
  &=& \frac{e^{i(\omega t+\phi)}}{2i}. \label{N1}
\end{eqnarray}
The above formula has used the approximation $\exp(x)\approx 1+x,~x\ll 1$ in line (\ref{A1}), and replaced the summation by integral in line (\ref{A2}), which are reasonable when $\tau/T\ll 1$. Following the same way, we prove that for $n=-1$
\begin{equation}\label{N2}
  \sum_{m=1}^{M}\frac{\xi w^{(m)}}{T} e^{i\omega t^{(m)}}e^{-i\omega t}
  \approx - \frac{e^{-i(\omega t + \phi)}}{2i}, 
\end{equation}
and for $n\neq\pm(lM\pm 1),~l=1,2,\cdots$
\begin{equation}\label{N3}
  \sum_{m=1}^{M}\frac{\xi w^{(m)}}{T}e^{-in\omega t^{(m)}}
  e^{in\omega t} \approx 0.
\end{equation}

Combining Eq. (\ref{N0}) and Eq. (\ref{N1}-\ref{N3}), the rectangular pulse sequence $s(t)$ can be written as
\begin{equation}
	s(t) = \sin \left( \omega t + \phi \right) + \xi\cdot f\left[ (lM \pm1)\omega \right].
\end{equation}
where $f\left[ (lM \pm 1)\omega \right]$ is an unknown function oscillating at the frequencies $(lM \pm 1)\omega$. Below the lowest frequency threshold $\Omega=(M-1)\omega$, the pulse sequence $s(t)$ has the same frequency components with the sinuous function $\sin{(\omega t+\phi)}$. Above the threshold, there are differences rapidly oscillating at frequency $(lM \pm 1)\omega,~l=1,2,\cdots$, with small amplitudes proportional to the pulse amplitude $\xi$. This means that one can always adjust the pulse number $M$ to guarantee a satisfactory frequency cutoff $\Omega=(M-1)\omega$, within which the pulse sequence $s(t)$ can be seen as identical to a sinuous function.

For an arbitrary real function $u(t)$ with finite frequency bandwidth, it can be equivalently written as
\begin{equation}\label{FU}
  u(t)=\frac{1}{\pi}\int_{\omega_{min}}^{\omega_{max}}d\omega
  A(\omega)\sin{(\omega t+\phi(\omega))}.
\end{equation}
where
\begin{eqnarray}
  A(\omega)&=&|\mathcal{U}(\omega)|,~
  {\phi(\omega)}=\pi-\arg{\{\mathcal{U}(\omega)\}},\\
  \mathcal{U}(\omega)&=&\int_{-\infty}^{+\infty}dt e^{-i\omega t}u(t).
\end{eqnarray}
In order to ensure a minimum frequency cutoff $\Omega$ for all the frequency terms, we set the pulse number $M=\Omega/\omega_{\rm min}+1$. In every interval $[t_{-}^{(m)},t_{+}^{(m)}]$, we suppose the pulse width $w^{(m)}(\omega)$ for all the frequency terms are the same and adjust the pulse amplitude $\xi(\omega)$ as follows 
\begin{equation}\label{A_F1}
  \xi(\omega)
  =\frac{A(\omega)}{w^{(m)}}\int_{t_{-}^{(m)}}^{t_{+}^{(m)}}dt\sin{(\omega t+\phi(\omega))}.
\end{equation}
Since the absolute value of $w_k^{(m)}$ is still unknown, we add a extra constraint that
\begin{equation}\label{A_F2}
  \frac{1}{\pi}\int_{\omega_{min}}^{\omega_{max}} d\omega\xi(\omega)=\xi,
\end{equation}
where $\xi$ is a chosen parameter. Combine Eq. (\ref{A_F1}) and (\ref{A_F2}), we obtain the pulse width for the $m$-th pulses in the sequence
\begin{equation}\label{TP2}
  w^{(m)}=\xi^{-1}\int_{t_{-}^{(m)}}^{t_{+}^{(m)}}dtu(t),
\end{equation}
which indicates that the pulse sequence $s(t)$ is only determined by the integral area of the time-dependent function $u(t)$ in every small time interval $\tau$.

\section{Error analysis of the PWM approximation}\label{APP2}
To analyze the accuracy of the PWM approximation in a more quantitive form, we use \emph{Taylor} series to expand Eq. (\ref{UP}) to the second order
\begin{eqnarray}\label{UP_TAY}
  U_{\rm PWM}^{(m)} &&=1+(-i)\sum_{k=0}^{K} \xi_k H_{k}w_k^{(m)}
  +\frac{(-i)^2}{2!}\sum_{k_1=0}^{K}\sum_{k_2=0}^{K} 
  \xi_{k_1}\xi_{k_2}H_{k_1}H_{k_2} t_{k_1}^{(m)}t_{k_2}^{(m)}
  +\mathcal{O}(\tau^3).
\end{eqnarray}
On the other hand, we use \emph{Dyson} series to expand the time-ordered integral Eq. (\ref{UACC}) to the second order
\begin{eqnarray}\label{U_TAY}
  \nonumber U^{(m)}
  &=&1+(-i)\sum_{k=0}^{K} H_{k}\int_{t_{-}^{(m)}}^{t_{+}^{(m)}}dtu_{k}(t) \\
  \nonumber && +(-i)^2\sum_{k_1=0}^{K}\sum_{k_2=0}^{K} H_{k_1} H_{k_2} 
  \int_{t_{-}^{(m)}}^{t_{+}^{(m)}}\int_{t_{-}^{(m)}}^{t_1}dt_1dt_2
  {u_{k_1}(t_1)u_{k_2}(t_2)} +\mathcal{O}(\tau^3).
\end{eqnarray}
In addition, the double integral in Eq. (\ref{U_TAY}) can be approximated by expanding the time-dependent variables $u_k(t)$ through \emph{Taylor} series, i.e.,
\begin{equation}\label{u_TAY}
  u_k(t)=u_k(t')+u'_k(t')\tau+\mathcal{O}(\tau^2),~t'\in[t_{-}^{(m)},t_{+}^{(m)}],
\end{equation}
Thus, according to Eq. (\ref{WID}) and Eq. (\ref{UP_TAY}-\ref{u_TAY}), we obtain the error formula for the PWM propagator \begin{equation}\label{SOD}
	U^{(m)} = U_{\rm PWM}^{(m)} + \mathcal{O}(\tau^3),
\end{equation}
which indicates that the PWM sequence has $2$nd order accuracy in simulating the control effect of a continuous control field. In other words, only by turning on/off the interactions, we can simulate the evolution of an arbitrary time-dependent system to the second order accuracy.

\section{Derivation of the higher-order PWM approximation}\label{APP3}
According to the error formula of the PWM type propagator Eq. (\ref{SOD}), one can explicitly write the exact propagator as \cite{Bandrauk1992, *Bandrauk1993}
\begin{equation}\label{H_UP}
  U^{(m)}=S_{2}^{(m)}[\tau] + \tau^3 C_3^{(m)}
  + \mathcal{O}(\tau^4),
\end{equation}
where $S_2^{(m)}[\tau] = U_{\rm PWM}^{(m)}$ represents the propagator in the interval $\tau$ under PWM approximation. On the other hand, the exact evolution $U^{(m)}$ can always be decomposed as
\begin{eqnarray}\label{H_U}
  \nonumber U^{(m)}&=& U\left( t_{+}^{(m)}, t_{+}^{(m)} - s\tau \right) 
  U\left( t_{+}^{(m)}-s\tau, t_{-}^{(m)}+s\tau \right)
  U\left( t_{-}^{(m)}+ s\tau, t_{-}^{(m)} \right), 
\end{eqnarray}
where $U(t_2,t_1)$ is the propagator from $t=t_1$ to $t=t_2$, $s$ is a real parameter to be determined. Substitute Eq. (\ref{H_UP}) into Eq. (\ref{H_U}) and choose $2s^3+(1-2s)^3=0$, one can prove that 
\begin{equation}
  U^{(m)}= S_2^{(m)}[s\tau] S_2^{(m)}[(1-2s)\tau] S_2^{(m)}[s\tau]
  +\mathcal{O}(\tau^4).
\end{equation}
This reveals that three PWM propagators can be concatenated to a higher-order form. To explore the result further, we continue to write the propagator explicitly as follows
\begin{equation}\label{C4}
  U^{(m)}=S_2^{(m)}[s\tau] S_2^{(m)}[(1-2s)\tau] S_2^{(m)}[s\tau] + \tau^4 C_4^{(m)} + \mathcal{O}(\tau^5).
\end{equation}
Using the symmetry of the two propagators
\begin{eqnarray}
	&U(t_{+}^{(m)},t_{-}^{(m)}) U(t_{-}^{(m)},t_{+}^{(m)}) = 1, \\
	&S_3^{(m)}[\tau] S_3^{(m)}[-\tau] = 1,
\end{eqnarray}
we immediately find that $C_4^{(m)}$ in Eq. (\ref{C4}) should be zero. In other words, three PWM propagators raise the approximation precision to the $4$th order. Define $S_{2n}^{(m)}[\tau]$ as the $2n$th order propagation scheme in small interval $\tau$, we obtain the higher-order PWM method as follows
\begin{equation}
  U^{(m)}=S_{2n}^{(m)}[\tau]+\mathcal{O}(\tau^{2n+1}),
\end{equation}
where 
\begin{eqnarray}
	S_{2n}^{(m)}[\tau]&=& S_{2n-2}^{(m)}[s\tau] S_{2n-2}^{(m)}[(1-2s)\tau]
  S_{2n-2}^{(m)}[s\tau], \\
  S_2^{(m)}[\tau] &=& U_{\rm PWM}^{(m)},
\end{eqnarray}
and
\begin{equation}
	s = {1}/{\left[ 2 + (-2)^{1/(2n-1)} \right]}.
\end{equation}

\section{Comparison between several short-time propagation methods}\label{A_CPR}
The strategy for solving time-dependent Schr\"odinger equation (TDSE) is to divide the total evolution operator into short segments in which the Hamiltonian does not change significantly \cite{Kosloff1988}
\begin{eqnarray}
	U(T,0) &=& \prod_{m=1}^{M} {U}^{(m)},\\
	U^{(m)}&=&\mathcal{T}
  \exp{\Big[-i\int_{(m-1)\tau}^{m\tau}dt H(t)\Big]},
\end{eqnarray}
where $\mathcal{T}$ is time-order interval, $\tau= {T}/{M}$ is an infinite small time interval. Consider the propagation operator in the $m$-th interval, it can be approximated by the Piecewise Constant  Scheme (PWC) 
\begin{equation}
  U_{\rm PWC}^{(m)}=\exp\left[ -i\tau H(t')\right],~t'\in[t_{-}^{(m)},t_{+}^{(m)}],
\end{equation}
or the Split Operator Scheme (SPO) (, also called the \emph{Suzuki-Trotter} Scheme)
\begin{equation}
	U_{\rm{SPO}}^{(m)}=\prod_{k=0}^{K}\prod_{k=K}^{0}\exp\left[ -i\tau/2 H_k \right].
\end{equation}
Approximation accuracy of the two schemes can be derived following the same procedure in Appendix \ref{APP2}. That is, we use $Taylor$ series to expand $U_{\rm{PWC}}^{(m)}$ and $U_{\rm{SPO}}^{(m)}$ to the second order, and compare the results with the second-order $Dyson$ series of the exact propagator $U^{(m)}$. We can prove that both PWC and SPO methods have $2$nd order accuracy in solving TDSE, which is the same as the PWM approximation. However, approximation error of these methods are slightly different in the frequency domain. \emph{Fourier} transform shows that, above the frequency threshold $\Omega$ PWC has smallest high-frequency noise among the three schemes; PWM is better than SPO for that the amplitudes of the noise is proportional to pulse amplitude $\xi$ (, as we said in Appendix \ref{APP}), while SPO corresponds to the special case in which $\xi\rightarrow\infty$ (see Sec. II). In conclusion, PWC is slightly more accurate than PWM than SPO, while they share the same order of approximation accuracy (, $2$nd order accuracy).

Computational complexity of the three schemes can be compared as follows. As defined in Sec. III, computational complexity $\mathcal{C}$ represents the number of multiplications required when calculating the short-time propagator. Expend the exponential functions into $Taylor$ series, i.e., $\exp[x]=1+x+\cdots+\mathcal{O}(x^{p+1})$, one can prove that $\mathcal{C}_{\rm PWC}=(p-1)N^3+KN^2$, where $N$ is the dimension of the system. In the same way and according to Eq. (\ref{UP_N}), we obtain that $\mathcal{C}_{\rm PWC}=(2K-1)N^3+(2K-1)N^2 + (p-1)KN$ (, $\mathcal{C}_{\rm SPO}$ is the same as $\mathcal{C}_{\rm PWC}$). Coefficient of the leading term of the three schemes are $p-1$ and $K-1$, which shows that PWM (, and SPO) saves computational complexity when the number of control variables is relatively small, or when the required numerical precision $p$ is very large. Consider other terms in the expression of $\mathcal{C}_{\rm PWM}$ (, and $\mathcal{C}_{\rm SPO}$), numerical simulation shows that a large dimension $N$ also out-stands the efficiency of PWM (, and SPO).

\bibliography{RPWM_4}
\end{document}